\documentclass[11pt,a4paper]{article}
\pdfoutput=1
\usepackage{jheppub}	

\usepackage{graphicx}

\def\lsim{\raise0.3ex\hbox{$\;<$\kern-0.75em\raise-1.1ex
\hbox{$\sim\;$}}}
\def\gsim{\raise0.3ex\hbox{$\;>$\kern-0.75em\raise-1.1ex
\hbox{$\sim\;$}}}

\def\be{\begin{equation}}
\def\ee{\end{equation}}
\def\ba{\begin{eqnarray}}
\def\ea{\end{eqnarray}}

\title{Can New Colored Particles Illuminate the Higgs?}

\author[a]{E.~Bertuzzo}
\author[a,b]{P.~A.~N.~Machado}
\author[a,b]{R.~Zukanovich Funchal} 

\emailAdd{enrico.bertuzzo@cea.fr}
\emailAdd{accioly@if.usp.br}
\emailAdd{zukanov@if.usp.br} 

\affiliation[a]{Institut de Physique Th\'eorique, CEA-Saclay, 91191 Gif-sur-Yvette, France}
 \affiliation[b]{
Instituto de F\'{\i}sica, Universidade de S\~ao Paulo, 
 C.\ P.\ 66.318, 05315-970 S\~ao Paulo, Brazil}

\abstract{ 
We analyze the behavior of Higgs to diphoton rate and Higgs
gluon-gluon production cross section in minimal extensions of the
Standard Model comprising new colored vector-like fermions that do not mix
with the ordinary ones. We compare these information with constraints coming 
from electroweak precision measurements. We compute pair production cross
sections for the lightest fermion and discuss the LHC bounds. Finally,
we study the phenomenology of possible quarkonium states composed by these new colored fermions.
} 

\begin{document}

\maketitle
\graphicspath{{plot-png/}}

\section{Introduction}
\label{sec:intro}

The auspicious discovery of a Higgs-like particle reported by
ATLAS~\cite{:2012gk} and CMS~\cite{:2012gu} experiments at the Large Hadron
Collider (LHC) marks the dawn of a new chapter in the field of
particle physics. This new particle seems to have many of the expected
properties of the long awaited Standard Model (SM) Higgs. However,
there are some alluring hints it might harbor new phenomena.

Analyses~\cite{Corbett:2012dm,Giardino:2012dp, Espinosa:2012im} of
ATLAS and CMS data at 7 and 8 TeV~\cite{:2012an,Chatrchyan:2012tx,
  cmspashig12015,cmspashig12020, atlasconf12091} combined with the
Tevatron experiments data~\cite{:2012zzl}, seem to indicate
discrepancies from the SM in the Higgs gluon-gluon production cross
section as well as in its diphoton decay. The former seems to be
$\sim$ 40\% smaller while the latter $\sim$ 2.5 larger than the SM
prediction.  While from the experimental point of view it is too early
to conclude that new physics is imperative, from the theory side
of view it is important to investigate what this might signify. In
particular, since in the last decades the SM has been submitted to
great experimental scrutiny passing all tests flawlessly it is not
inconceivable that the first sign of new physics that will be observed
at the LHC is manifested through loop effects.

In this paper we will assume this is the case and try to understand to
what extent one can explain an increase of the $H\to \gamma \gamma$
partial decay width as well as a decrease of the Higgs gluon-gluon
production cross section, by introducing the minimum number of extra
charged fermion states that couple to the Higgs.  Since the first
measurements by the LHC experiments suggested the observed Higgs to
diphoton rate was larger than the SM
one~\cite{Chatrchyan:2012tw,ATLAS:2012ad}, a number of theoretical
papers have attempted to explain this effect by introducing new
fermions~\cite{Kumar:2012ww,Dawson:2012di,Bonne:2012im,Carena:2012xa,
Joglekar:2012vc,ArkaniHamed:2012kq,Almeida:2012bq,Davoudiasl:2012ig,Voloshin:2012tv,Kearney:2012zi,Lee:2012wz}, 
scalars~\cite{Draper:2012xt,Kumar:2012ww,Akeroyd:2012ms,Carena:2012xa,Wang:2012ts,Bertolini:2012gu,Chun:2012jw,Chang:2012ta,
Kribs:2012kz,Dorsner:2012pp}, 
a spin-2 resonance~\cite{Urbano:2012tx}, 
in SUSY models~\cite{Cao:2011pg,Cao:2012fz,Carena:2011aa,Carena:2012gp,Sato:2012bf,Boudjema:2012in,Kitahara:2012pb,SchmidtHoberg:2012yy,Haba:2012zt,Bellazzini:2012mh}
or other scenarios~\cite{Alves:2011kc,Carmi:2012in,Kobakhidze:2012wb,Olesen:2012zb,Alves:2012ez}. 
Some consequences to the vacuum stability of a wrong-sign gluon-gluon
amplitude are discussed in \cite{Reece:2012gi}.

In a previous paper \cite{Almeida:2012bq} we have shown how it is
possible to increase the Higgs diphoton rate by the introduction of
new colorless fermions that do not mix with other SM fermions but mix
among themselves.  Our goal in this paper is to extend this idea to
include colored particles.  We will discuss the properties (masses,
charges, couplings) that these new fermions have to satisfy in order
to be at the same time in consonance with the currently observed $H\to
\gamma \gamma$ width and with the gluon-gluon Higgs production cross
section.  We will consider the extra fermions in their smallest
allowed $ \rm SU(2)_L \times U(1)_Y$ representations and assume that
the additional states that will accompany these fermions in a complete
model, are heavy enough to play no significant role in the Higgs
sector.  We will confront our results with the electroweak precision
tests and discuss the production of these new particles at the LHC.

This work is organized as follows. In the Sec.~\ref{sec:ggloop}
we discuss, in a generic way, how the Higgs diphoton rate and 
the Higgs production by gluon-gluon fusion can be affected by 
new charged fermions that receive part of their mass from 
the electroweak symmetry breaking mechanism. 
In particular, we focus on the correlated role they may play 
in $\sigma(gg\to h)$ and $\Gamma(h\to \gamma \gamma)$.  
In Sec.\ref{sec:framework} we describe the two simple frameworks  
we will introduce for new colored vector-like fermions that do not 
mix with the SM ones. They will involve the smallest 
$\rm SU(2)_L$ representations of these fields. 
In the doublet-singlet (triplet-doublet) model
we introduce two new fields: a $\rm SU(2)_L$ doublet (triplet) 
and a singlet (doublet).  From the color point of view  
we will assume these new field can be either
$\rm SU(3)_c$ triplets, sextets or octets.
Given a fixed gauge group representation we will discuss 
in Sec.~\ref{sec:analysis}
how these new particles can impact $\sigma(gg\to h)$, 
$\Gamma(h\to \gamma \gamma)$ 
and the $S,T,U$ oblique parameters, as a function of their 
masses and couplings. This aims to outline what are the 
allowed properties of these colored fermion fields 
that might be needed to explain future LHC data.
In Sec.\ref{sec:signals} we discuss the direct pair 
production of these particles at the LHC as well as 
the possible formation of quarkonia.  
Finally in Sec. \ref{sec:conclusions} we draw our conclusions.

\section{Effects of New Fermions in the Higgs Production and 
Diphoton Decay}
\label{sec:ggloop}

It may be fortuitous that the main production process of this new
particle at the LHC is the gluon-gluon fusion and the most conspicuous
signal observed up to now comes from its diphoton decay. Not only both
happen at the loop level, probing quantum corrections coming from any
new states carrying color or charge, moreover, these corrections are
correlated for colored states.

We will consider here that the 125 GeV particle discovered at the LHC is in
fact the SM Higgs boson, $h$, a fundamental scalar transforming as part
of a $\rm SU(2)_L \times U(1)_Y$ doublet with the SM Higgs charge 
assignments and hypercharge Y=1/2. 
Assuming new colored fermions couple to the Higgs field, 
we can define the ratio between the production cross section 
including these new colored particles, $\sigma(g g \to h)$, and the 
SM one, $\sigma (gg \to h)\vert_{\rm SM}$, as 

\begin{equation}
R_{\rm GG} \equiv 
\frac{\sigma(gg \to h)}{\sigma (gg \to h)\vert_{\rm SM}} 
\approx \Big \vert 1 - 9.46 \, \sum_i \frac{I_i(R)}{Q_i^2 D_i(R)} \,\delta_c^i 
\Big \vert^2 \, ,
\label{eq:GG}
\end{equation}
where $I_i(R)$ and $D_i(R)$ are, respectively, the index and dimension of
the $\rm SU(3)_c$ representation of these new fermions, $Q_i$ their
electric charge and $\delta_c^i$ parametrizes their contributions to the
$hGG$ vertex and is related to the $h \to \gamma \gamma$ rate as will
be discussed below. Here we have only considered the top quark in
estimating the SM cross section.

\begin{figure*}[tb]
\begin{center}
 \includegraphics[width=0.49\textwidth]{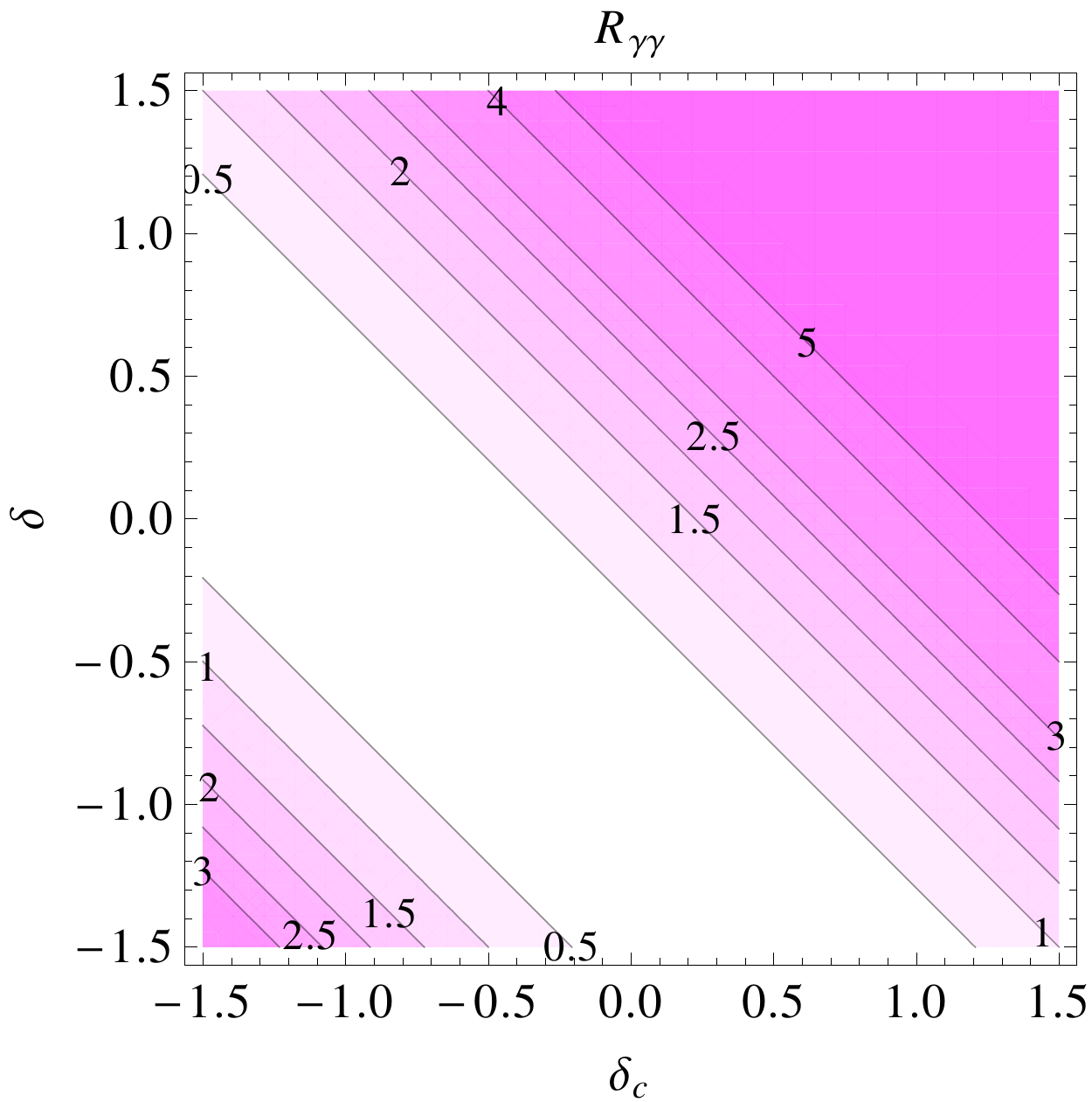}
\vspace{-2mm}
\end{center}
\vspace{-0.1cm}
\caption{Iso-contours  of the ratio $R_{\gamma \gamma}$ in the plane 
$\delta_c$ versus $\delta$.}
\label{fig:gg-gen}
\end{figure*}

Allowing for new charged uncolored fermions, as well as new colored ones, to 
couple to $h$, we can also define the ratio between the $h \to 
\gamma \gamma$ rate including this new states, $\Gamma(h \to \gamma \gamma)$,
and the SM one, $\Gamma(h \to \gamma \gamma)\vert_{\rm SM}$, as 

\begin{equation}
R_{\rm \gamma \gamma} \equiv 
\frac{\Gamma(h\to \gamma \gamma)}{\Gamma (h \to \gamma \gamma)\vert_{\rm SM}} 
\approx \Big \vert 1 + \sum_i (\delta^i + \delta_c^i) \Big \vert^2 \, ,
\label{eq:gammagamma}
\end{equation}
where $\delta^i$ and $\delta_c^i$ parametrize, respectively, the uncolored
and colored particles new contributions with respect to the SM. To a
very good approximation~\cite{Ellis:1975ap}, for new fermion masses
greater or of the order of the Higgs mass

\begin{eqnarray}
\delta^i &=& \frac{(Q^{\rm u}_i)^2}{A_{\rm SM}^{\gamma \gamma}} \, \frac{v}{2}  \frac{d \ln M^{\rm u \, 2}_i(v)}{dv} \frac{4}{3}(1+ \frac{7}{120}\frac{M_h^2}{M^{\rm u \, 2}_i}) \, ,\\
\delta_c^i &=& \frac{Q^2_i \, D_i(R)}{A_{\rm SM}^{\gamma \gamma}} \frac{v}{2} \frac{d \ln M_i^2(v)}{dv} \frac{4}{3}(1+ \frac{7}{120}\frac{M_h^2}{M_i^2}) \, ,
\label{eq:deltas}
\end{eqnarray}
where $M_h$ is the Higgs boson mass, 
$v$ is the Higgs vacuum expectation value, $A_{\rm SM}^{\gamma \gamma}$ is
the main contribution to the SM Higgs to diphoton amplitude, namely,
the one coming from the W boson and the top, {\it i.e.}, $A_{\rm
  SM}^{\gamma \gamma}=-6.48$.  Here $Q^{\rm u}_i$ ($Q_i$) and $M^{\rm u}_i$ ($M_i$)
are the electric charge and masses of the new uncolored (colored) fermions.
As discussed in  many papers, fermion mixing allows for the new terms to 
have the opposite sign of the top one. This can, as we will see, 
increase the Higgs to diphoton rate and at the same time decrease 
the Higgs production through gluon-gluon fusion.

In Fig.~\ref{fig:gg-gen} we show iso-contour lines of constant
$R_{\gamma \gamma}$ in the plane $\delta_c$ versus $\delta$ (both assumed real).  The SM
corresponds to the point $\delta=\delta_c=0$ in this plot.  In
Ref.~\cite{Almeida:2012bq} we have discussed how uncolored fermions
can affect this ratio when they mix among themselves but do not mix
with SM ones. This case corresponds to values of $\delta$ for which
$\delta_c=0$, so to increase the $R_{\gamma \gamma}$  ratio only 
positive values of $\delta$ would be allowed. However, if one 
also allows for a  non zero $\delta_c$, there are  more possibilities
to explain the same $R_{\gamma \gamma}$  ratio.

\begin{figure*}[tb]
\begin{center}
 \includegraphics[width=0.3\textwidth]{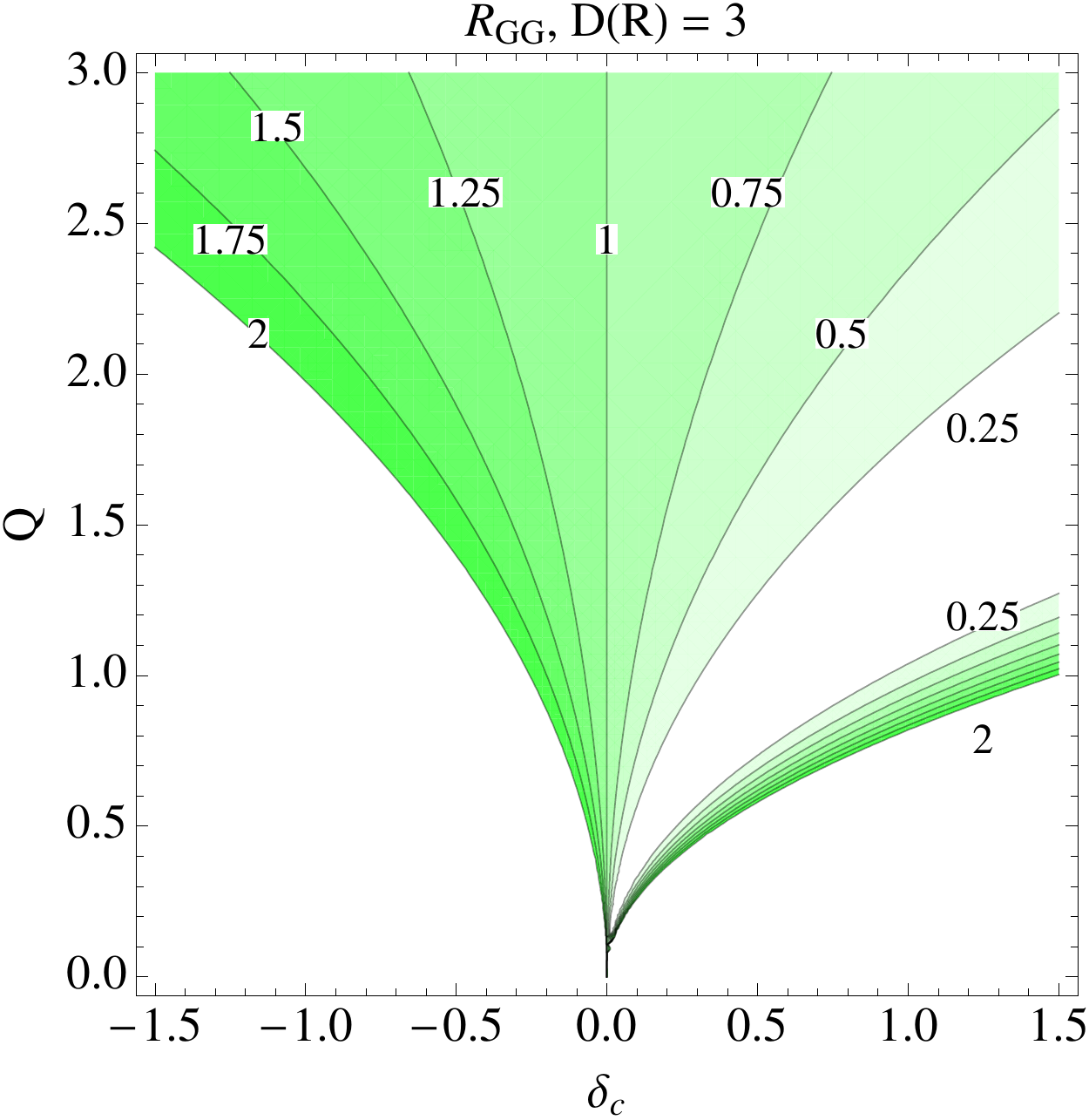}
 \includegraphics[width=0.3\textwidth]{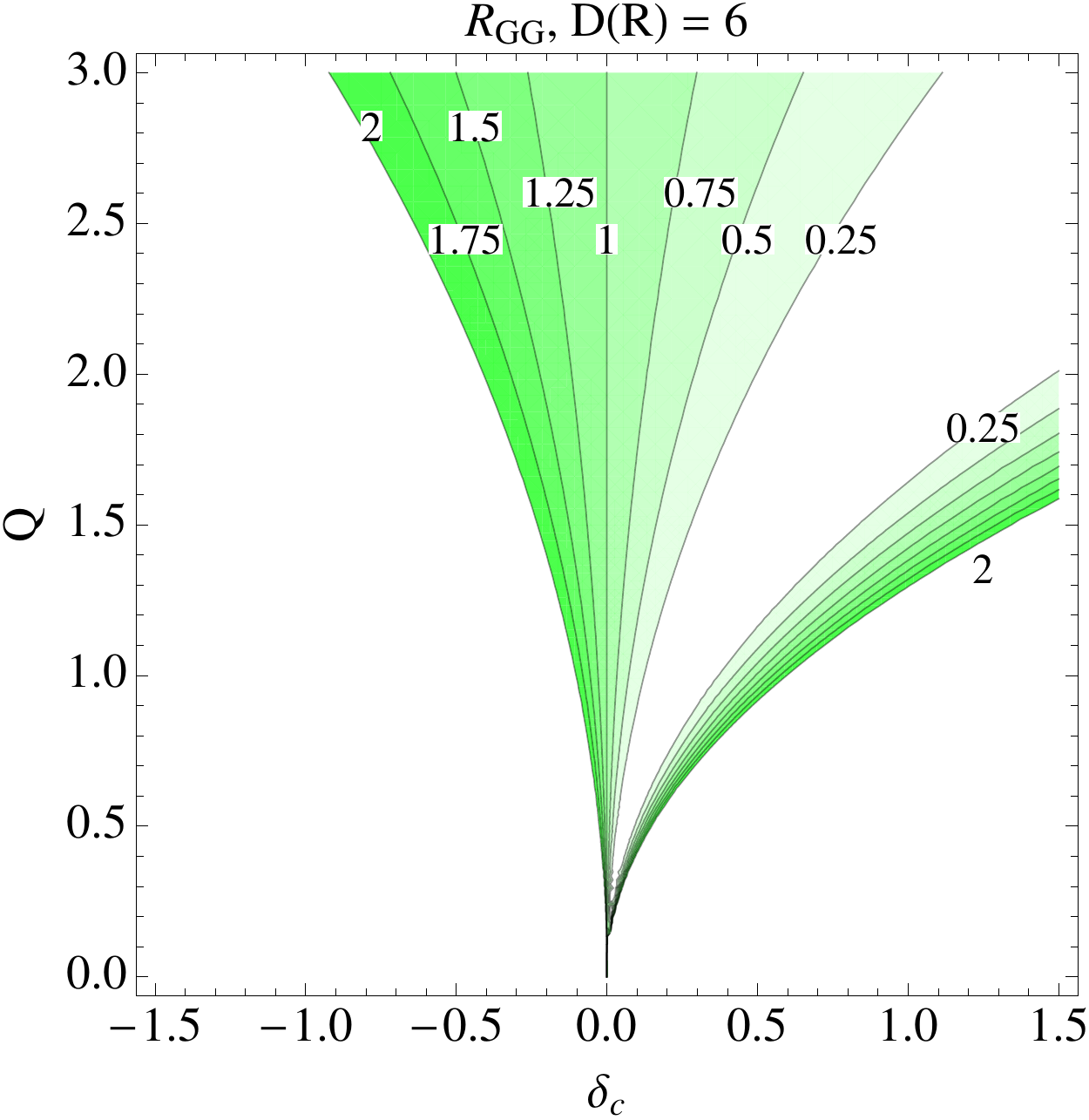}
 \includegraphics[width=0.3\textwidth]{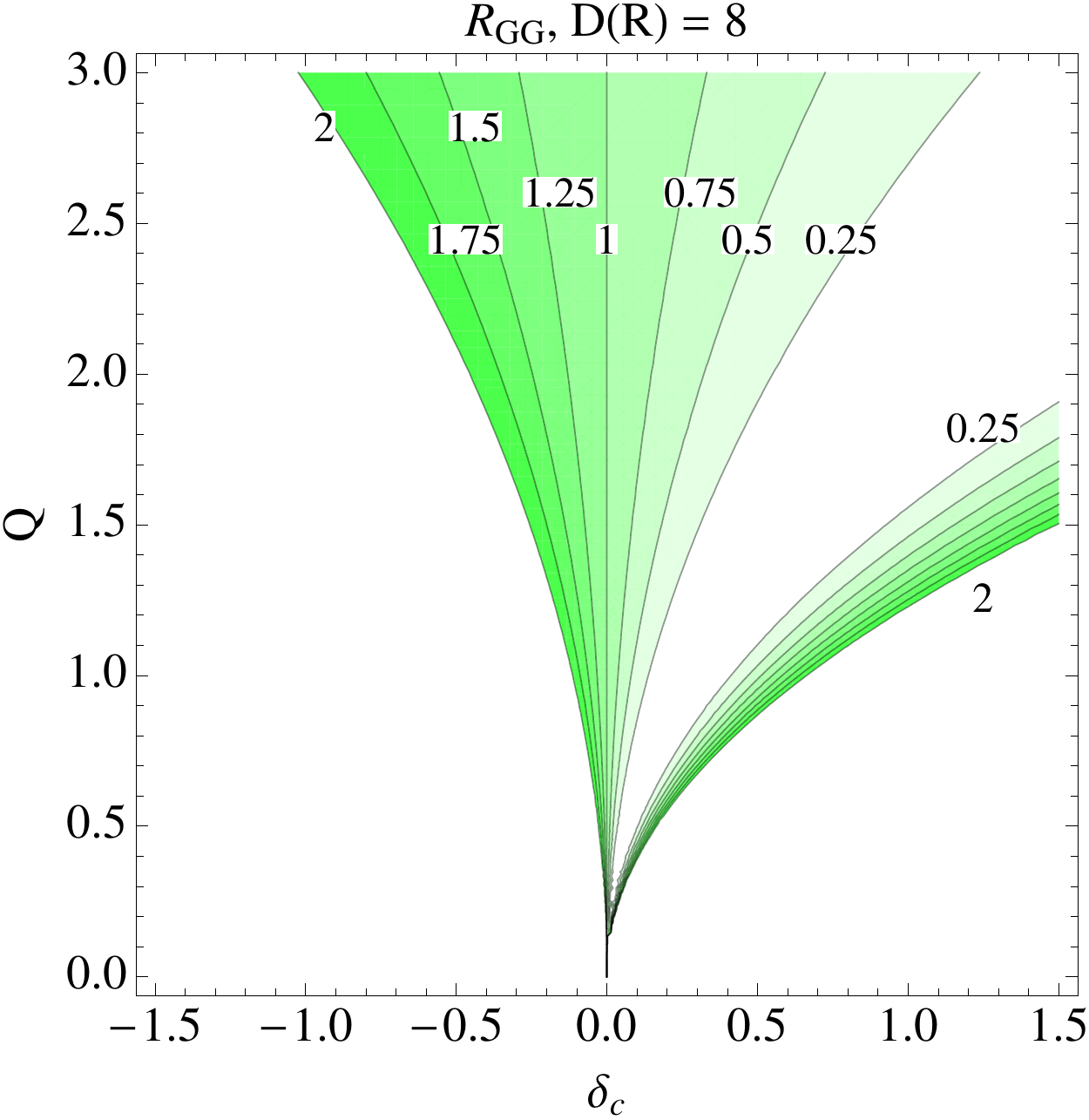}
\vspace{-2mm}
\end{center}
\vspace{-0.1cm}
\caption{Iso-contours  of the ratio $R_{GG}$ in the plane 
$\delta_c$ versus $Q$, for different $\rm SU(3)_c$ representations: 
triplets (left panel), sextets (middle panel) and octets (right panel).}
\label{fig:gluglu-gen}
\end{figure*}

In Fig.~\ref{fig:gluglu-gen} we show how the ratio $R_{GG}$ changes as
a function of $\delta_c$ and the charge $Q$ assuming all the new particles to have common 
electric charge and color representation.  This figure along with Fig.~\ref{fig:gg-gen} can be
used to understand what kind of new fermions can enter the loop
diagrams in order to explain the experimental data.  For instance, if
we want to get a 50\% decrease of the gluon-gluon production cross
section with respect to the SM one, we see what are the combinations
of $\delta_c$ and $Q$ that are allowed in all three color
representations.  Once we select a $\delta_c$, one can go back to
Fig.~\ref{fig:gg-gen} to see what values of $\delta$ are needed to
explain $R_{\gamma \gamma}$.  We see from these figures that to
explain $R_{\gamma \gamma} \sim 2$ with only colored fermions, $\delta_c
\sim 0.5$ is needed. This can be, in principle, achieved, in any of
the three color representations, even if $R_{GG}$ deviates from the SM 
expectation by $\sim$ 10\%.

After this general discussion about the interplay between the effect
of new colored fermions in $R_{GG}$ and $R_{\gamma \gamma}$ let us now
discuss some specific realizations.

\section{Brief Description of Two Simple Frameworks}
\label{sec:framework}

As the addition of a sequential fourth fermion generation to the particle 
content of the SM is now ruled out by the LHC data at the 5$\sigma$ level~\cite{Eberhardt:2012gv}, 
we will assume the new colored fermions to be vector-like.
Moverover, these new particles will not mix with the SM fermions, 
but will couple to the Higgs and the gauge sector respecting the SM symmetry
group.  One can envisage a new quantum number in connection to an unbroken or 
nearly unbroken symmetry, exclusive to the new sector, in order to 
forbid this mixing~\cite{Eboli:2011hr}.

In order to be able to build a renormalizable coupling term with the
SM Higgs field and to have mixing we need to introduce at least two
extra fermion fields. 
As in Ref.~\cite{Almeida:2012bq} we will consider only the
smallest allowed $\rm SU(2)_L \times U(1)_Y$ representations of the new 
fermion states, \emph{i.e.}  singlets, doublets and triplets. 
We will also allow for these new colored fermions to be either 
$\rm SU(3)_c$ triplets, sextets or octets. 
The choice of the $\rm SU(3)_c \times SU(2)_L \times U(1)_Y$
representation of the new fermions will fix their couplings to the SM 
gauge sector. Given that, the only free parameters will be their couplings
to the Higgs, their charges and their masses.  

\subsection{\bf Doublet-singlet model (2+1)}

 The lagrangian describing the new fermion masses and couplings 
with the Higgs  is 

\begin{equation}
{-\cal L_{\rm H}^{\rm 2+1}} = c \, \overline{S}_R \, H D_L + c
\,\tilde{H} \, \overline{D}_R P_L  S_L + m_1 \overline{D}_R P_L D_L +
m_2 \overline{S}_R P_L S_L  + \rm h.c.,
\label{eq:h-2+1}
\end{equation}
where $S$ and $D$ are, respectively, a $\rm SU(2)_L$ singlet and 
doublet field, $\tilde{H}= i \tau_2 H^{*}$, $c$ is the Yukawa coupling to the 
Higgs, $P_{L,R} = \frac{1}{2}(1\pm \gamma_5)$,
$m_{1,2}$ are the vector-like $S,D$ masses.

As usual the gauge interactions with the SM fields are described by the 
coupling with the SM fields and introduced via covariant derivatives~\cite{Almeida:2012bq} 
\begin{equation}
{\cal L_{\rm I}^{\rm 2+1}} = 
i \overline S \gamma^\mu \left( \partial_\mu -i g'y \, B_\mu \right) S + i \overline D \gamma^\mu 
\left( \partial_\mu - i g \,W_\mu^a 
T^a -i g'\hat{y} \, B_\mu \right) D , 
\label{eq:gauge-2+1}
\end{equation}
where $g'=e/c_W$, $g=e/s_W$ are the SM couplings, $s_W^2 = \sin^2\theta_W = 1- c_W^2 \equiv 1- M_W^2/M_Z^2$, 
(with $M_Z$ and $M_W$, respectively, the Z and W boson masses), and $y$ ($\hat y = y
-1/2$) is the hypercharge of the singlet (doublet).

\subsection{\bf Triplet-doublet model (3+2)}

We will consider the following mass lagrangian for the new states
\begin{equation}
{-\cal L_{\rm mass}^{\rm 3+2}} = c \left ( \overline{D}_R \, T_L H +
\overline{D}_L \, T_R H \right)   + m_1 \, \overline T_L T_R + m_2 \, \overline D_L 
D_R + \rm h.c.   \, , 
\label{eq:hm-3+2}
\end{equation}
where $D$ and $T$ are, respectively, a $\rm SU(2)_L$ doublet and triplet fields,
$c$ is their coupling to the SM Higgs field and $m_1$ and $m_2$ their 
vector-like masses. 

The gauge interactions with the SM fields are described again in the 
usual way by~\cite{Almeida:2012bq} 
\begin{equation}
{\cal L_{\rm I}^{\rm 3+2}} = 
i \overline D \gamma^\mu \left( \partial_\mu - i g \, W_\mu^a  
T^a -i g'y \, B_\mu \right) D + i \bar T \gamma^\mu 
\left( \partial_\mu - i g W_\mu^a 
T^a -i g'\hat{y} \, B_\mu \right) T \, ,
\label{eq:gauge-3+2}
\end{equation}
here $y$ ($\hat y = y -1/2$) is the hypercharge of the doublet (triplet). 

\subsection{\bf Color Interactions}

Independent of their $\rm SU(2)_L$ representations our colored 
fermions will couple to gluons according to their $\rm SU(3)_c$ 
representation. The strong sector lagrangian can be written as 
\begin{equation}
 {\cal L_{\rm S}} = \sum_i \overline \psi_i \, i \, \gamma^\mu\, 
(\partial_\mu - i g_s t^{A}G^{A}_\mu) \psi_i \, ,
\label{eq:strong}
\end{equation}
where $g_s$ is the strong coupling constant and the sum runs over 
all the mass eigenstates of the new fermions. 
These will be three states in the singlet-doublet model and 
five states in the triplet-doublet model.
The dependence on the color representation is encoded in 
the $t^{A}$ matrices.

\section{Analysis of the Viability of the Models}
\label{sec:analysis}

For each model we have investigated the possibility of the new 
fermions to be able to explain the current allowed regions of 
$R_{GG}$ and $R_{\gamma \gamma}$, according to Ref.~\cite{Corbett:2012dm}, 
and at the same type fulfill the requirements of the electroweak precision 
tests.

New colored states necessarily contribute to the 
vacuum polarization amplitudes of the electroweak gauge bosons $\Pi^{\mu \nu}_{ab}(q^2) =
- i g^{\mu \nu} \Pi_{ab}(q^2) + q^\mu q^\nu$
terms~\cite{Altarelli:1990zd,Peskin:1991sw}. 
The new  effects can be 
parametrized by the so-called quantum oblique parameters $S$, $T$ and 
$U$ defined as~\cite{Peskin:1991sw} 

\begin{eqnarray}
\alpha(M_Z^2) \, S^{\text{NP}} & = & \frac{4 s_W^2 c_W^2}{M_Z^2} \left [\Pi^{\text{NP}}_{ZZ} (M_Z^2) - \Pi^{\text{NP}}_{ZZ} (0)
    -\Pi^{\text{NP}}_{\gamma \gamma}(M_Z^2) - \frac{c_W^2-s_W^2}{c_W
        s_W} \, \Pi^{\text{NP}}_{\gamma Z}(M_Z^2)\right] \, ,\nonumber
      \\ 
\alpha(M_Z^2) \, T^{\text{NP}} & = & \frac{\Pi^{\text{NP}}_{WW}(0)}{M_W^2}
        - \frac{\Pi^{\text{NP}}_{ZZ}(0)}{M_Z^2} \, , \nonumber
          \\ 
\alpha(M_Z^2) \, U^{\text{NP}} & = & 4 s_W^2 \left [
            \frac{\Pi^{\text{NP}}_{WW}(M_W^2)-\Pi^{\text{NP}}_{WW}(0)}
                {M_W^2} - c_W^2 \left(
                \frac{\Pi^{\text{NP}}_{ZZ}(M_Z^2)-\Pi^{\text{NP}}_{ZZ}(0)}{M_Z^2}\right)
                    \nonumber \right .\\ & & \left . - 2 s_W c_W \,
                    \frac{\Pi^{\text{NP}}_{\gamma Z}(M_Z^2)}{M_Z^2} -
                       s_W^2 \, \frac{\Pi^{\text{NP}}_{\gamma
                            \gamma}(M_Z^2)}{M_Z^2} \right ],
\label{eq:STU}
\end{eqnarray}

The most recent electroweak study of the SM observables, finds the  following 
best fit values for the oblique parameters~\cite{Baak:2012kk}

\begin{eqnarray}
\Delta S & = & S - S_{\rm SM}  =  0.03 \pm 0.10 \nonumber \\
\Delta T & = & T - T_{\rm SM}  =  0.05 \pm 0.12 \nonumber \\
\Delta U & = & U - U_{\rm SM}  =  0.03 \pm 0.10 
\label{eq:stulimits}
\end{eqnarray}
for the reference Higgs and top masses $M_{H,\rm ref} = 126$ GeV 
and $m_{t,\rm ref}= 173$ GeV, with the associated correlation matrix 

\begin{equation}
V= \left( \begin{array}{ccc}
1 & + 0.89 & -0.54 \\
+0.89 & 1 & -0.83 \\
-0.54 & -0.83 & 1
\end{array}\right).
\label{eq:corr}
\end{equation}

As in Ref.~\cite{Almeida:2012bq}, we will include these constraints in our 
models by minimizing the $\chi^2$ function 

\begin{equation}
\chi^2 = \sum_{i,j} (X_i^{\text{NP}} -
X_i)(\sigma^2)^{-1}_{ij}(X_j^{\text{NP}} - X_j),
\label{eq:chi2}
\end{equation}
where $X_i= \Delta S, \Delta T, \Delta U$, are the fitted values of
the oblique parameters with their corresponding uncertainties
$\sigma_i$ defined in Eq.(\ref{eq:stulimits}), $X_i^{\text{NP}} =
S^{\text{NP}}, T^{\text{NP}}, U^{\text{NP}}$ are the contributions
from the extra states and $(\sigma_{ij})^2 \equiv \sigma_i V_{ij}\sigma_j$.  We
will allow the values of the parameters of our models to vary such
that $\Delta \chi^2 = (3.53, 7.81, 11.3)$, which correspond to (68\%,
95\%, 99\%) CL for a three-parameter fit.  
In a complete model there will be other states that appear in conjunction 
with the new  colored fermions we are considering in this work. 
In principle, these particles also contribute to the vacuum polarization 
amplitudes. Having said that, we will here  assume these other particles are 
sufficiently heavy to play no consequent role.

\begin{figure*}[tb]
\begin{center}
 \includegraphics[width=0.95\textwidth]{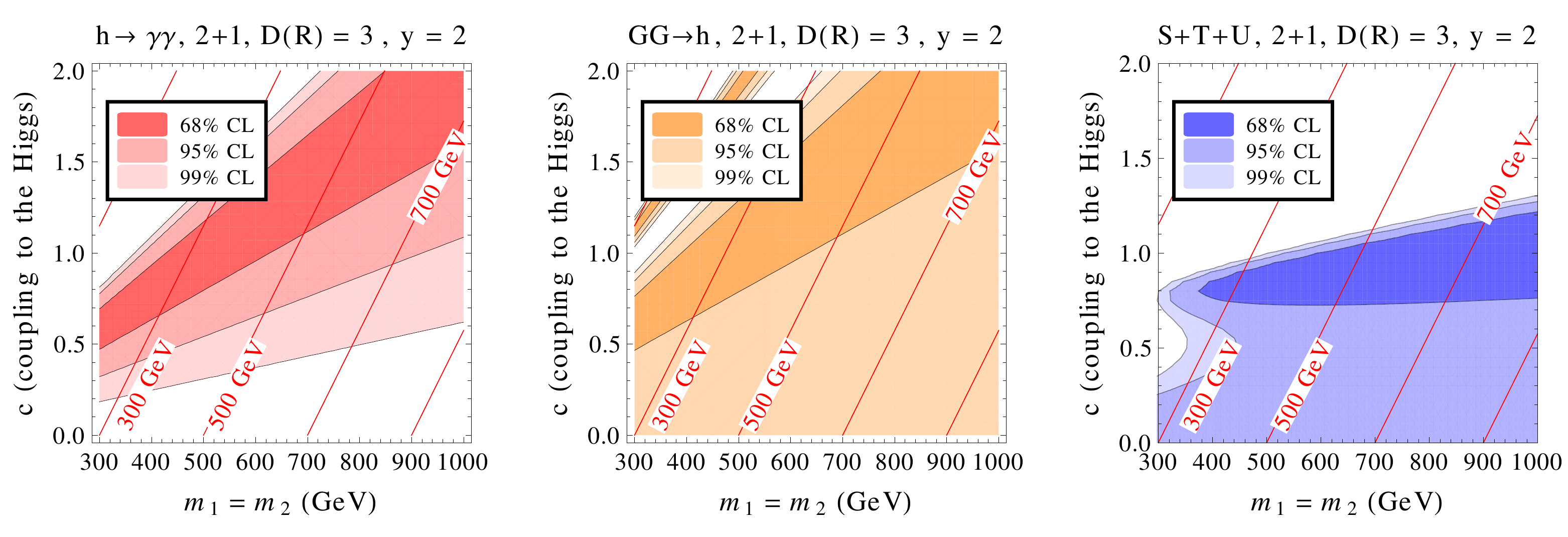}
\vspace{-2mm}
\end{center}
\vspace{-0.1cm}
\caption{ Allowed region in the plane $m_1=m_2$ and $c$ for the 2+1
  model with hypercharge $y=2$ and color triplet fermions (D(R)=3) for
  $R_{\gamma \gamma}$ (left panel), $R_{GG}$ (middle panel) and the
  oblique parameters $S,T$ and $U$ (right panel).  The allowed regions
  for $R_{\gamma \gamma}$ and $R_{GG}$ are taken from
  Ref.~\cite{Corbett:2012dm}. We also show the iso-lines of constant
  mass for the lightest colored fermion (in red). }
\label{fig:2+1gen}
\end{figure*}

\begin{figure*}[tb]
\begin{center}
 \includegraphics[width=0.95\textwidth]{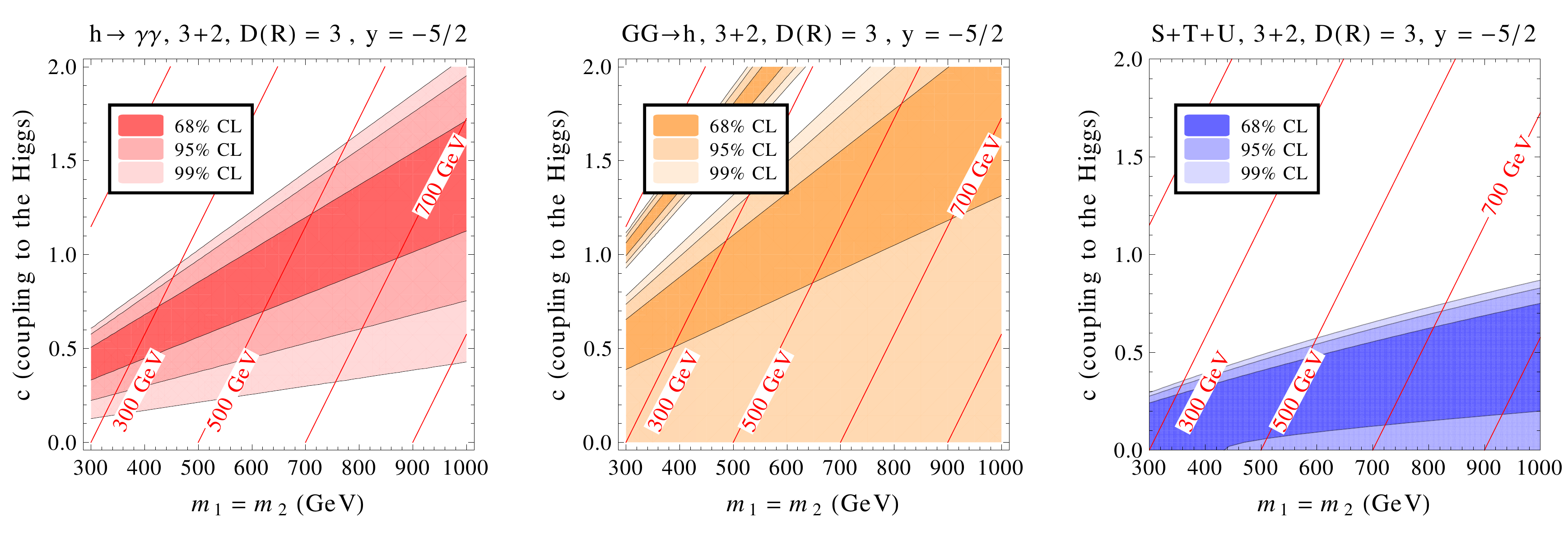}
\vspace{-2mm}
\end{center}
\vspace{-0.1cm}
\caption{ Allowed region in the plane $m_1=m_2$ and $c$ for the 3+2
  model with hypercharge $y=-5/2$ and color triplet fermions (D(R)=3)
  for $R_{\gamma \gamma}$ (left panel), $R_{GG}$ (middle panel) and
  the oblique parameters $S,T$ and $U$ (right panel).  The allowed
  regions for $R_{\gamma \gamma}$ and $R_{GG}$ are taken from
  Ref.~\cite{Corbett:2012dm}. We also show the iso-lines of constant
  mass for the lightest colored fermion (in red).  }
\label{fig:3+2gen}
\end{figure*}

To illustrate the behavior of $R_{\gamma \gamma}$, $R_{GG}$ and 
$S,T$ and $U$ combined in the models with colored fermions 
as a function of the parameters $m_1=m_2$ and $c$, in Fig.~\ref{fig:2+1gen} we show, 
in the case of triplet colored fermions in the 2+1 model with $y=2$, 
the allowed regions for $R_{\gamma \gamma}$ (left panel), 
$R_{GG}$ (middle panel) and $S,T$ and $U$ combined (right panel). 
We can observe that in this example it is possible to explain 
$R_{\gamma \gamma}$ and $R_{GG}$, even in the current 68\% CL allowed 
region and still be compatible with the oblique parameters.
In Fig.~\ref{fig:3+2gen}  we show the same three panels for the 
triplet colored fermions in the 3+2 model with $y=-5/2$.
Here the tension with the oblique parameters is apparent and mainly driven 
by $S$. In this case 
it is quite difficult to explain the experimental ratios and at the 
same time satisfy the electroweak tests, unless $R_{\gamma \gamma}$ and 
$R_{GG}$ turn out to be respectively lower and higher than the 
present best fit value, {\it i.e.} closer to their SM predictions.

\begin{figure*}[tb]
\begin{center}
 \includegraphics[width=0.3\textwidth]{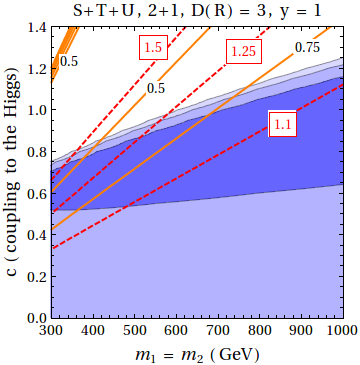}
 \includegraphics[width=0.3\textwidth]{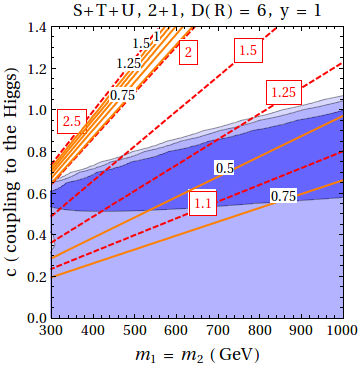}
 \includegraphics[width=0.3\textwidth]{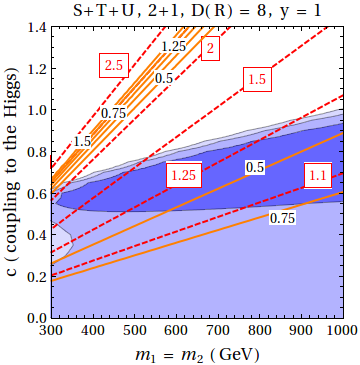}
\vglue 0.8mm 
\includegraphics[width=0.3\textwidth]{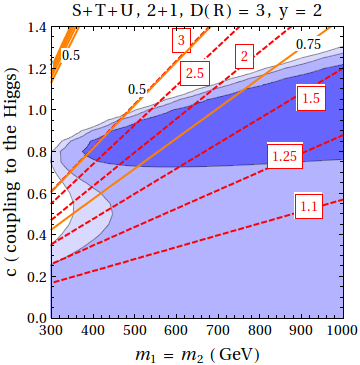}
 \includegraphics[width=0.3\textwidth]{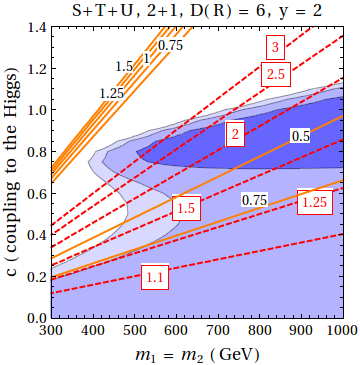}
 \includegraphics[width=0.3\textwidth]{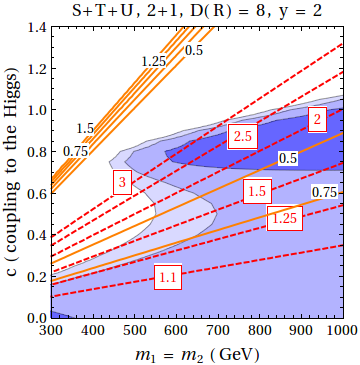}
\vglue 0.8mm 
\includegraphics[width=0.3\textwidth]{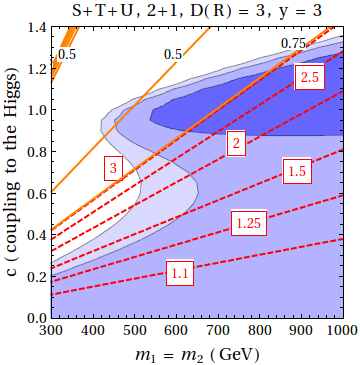}
 \includegraphics[width=0.3\textwidth]{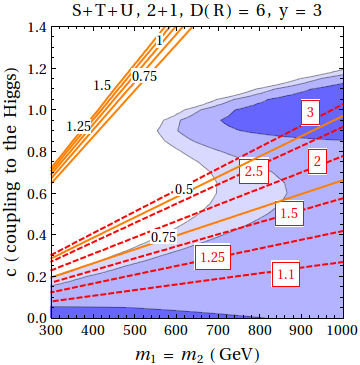}
 \includegraphics[width=0.3\textwidth]{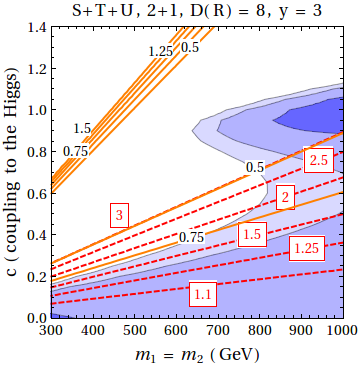}
\vspace{-2mm}
\end{center}
\vspace{-0.1cm}
\caption{Allowed regions in the plane $m_1=m_2$ and $c$ for combined $S,T$ and 
$U$ at 68\%, 95\% and 99\% CL in the 2+1 model.
Panels on the left are for SU(3)$_c$ triplets, on the middle for 
SU(3)$_c$ sextets and on the right for SU(3)$_c$ octets.
Panels from top to bottom are, respectively,  
 for $y=1,2$ and $3$.
We also show iso-lines of constant $R_{\gamma \gamma}$ (dashed red) and $R_{GG}$ (solid orange). 
}
\label{fig:2+1global}
\end{figure*}

\begin{figure*}[ptb]
\begin{center}
 \includegraphics[width=0.3\textwidth]{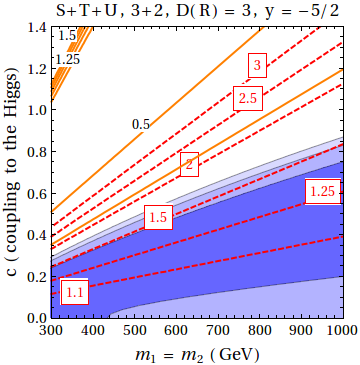}
 \includegraphics[width=0.3\textwidth]{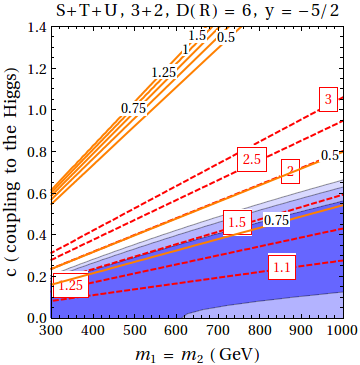}
 \includegraphics[width=0.3\textwidth]{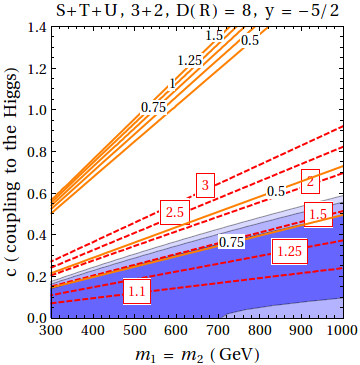}
\vglue 0.8mm 
 \includegraphics[width=0.3\textwidth]{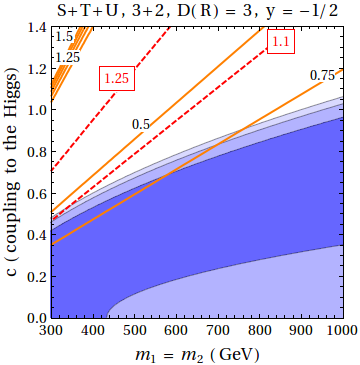}
 \includegraphics[width=0.3\textwidth]{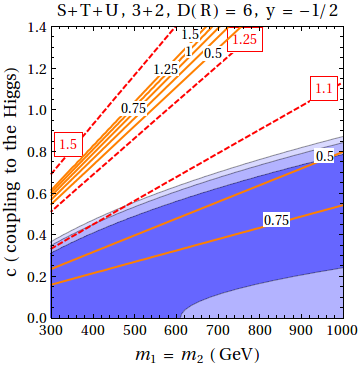}
 \includegraphics[width=0.3\textwidth]{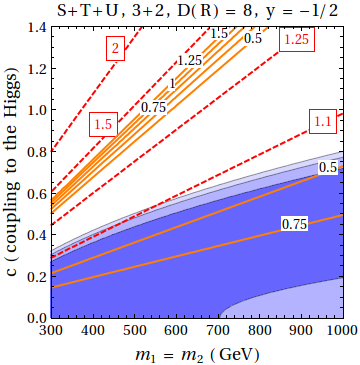}
\vglue 0.8mm 
\includegraphics[width=0.3\textwidth]{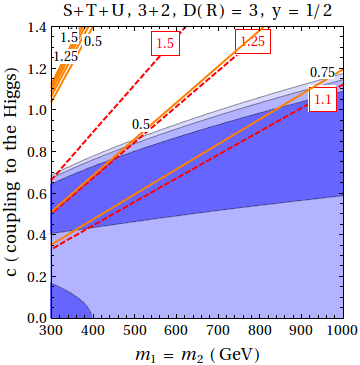}
 \includegraphics[width=0.3\textwidth]{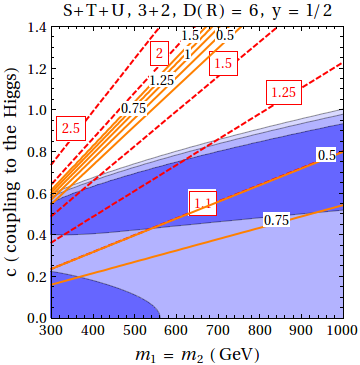}
 \includegraphics[width=0.3\textwidth]{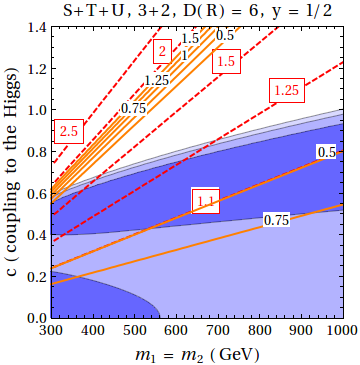}
\vglue 0.8mm 
\includegraphics[width=0.3\textwidth]{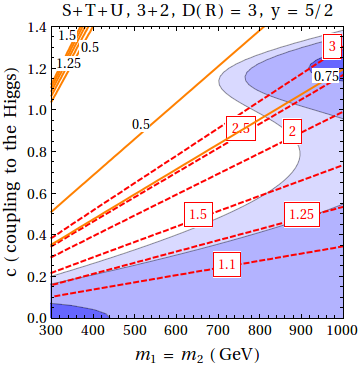}
 \includegraphics[width=0.3\textwidth]{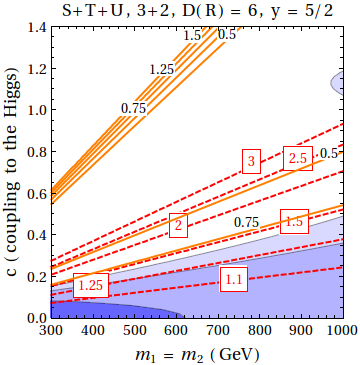}
 \includegraphics[width=0.3\textwidth]{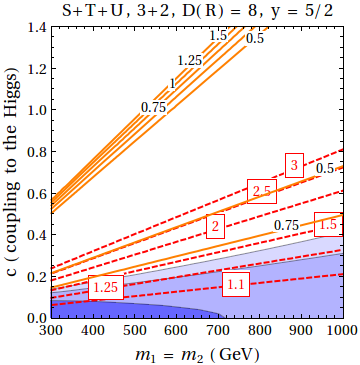}
\vspace{-2mm}
\end{center}
\vspace{-0.1cm}
\caption{Allowed regions in the plane $m_1=m_2$ and $c$ for combined $S,T$ and 
$U$ at 68\%, 95\% and 99\% CL in the 3+2 model.
Panels on the left are for SU(3)$_c$ triplets, on the middle for 
SU(3)$_c$ sextets and on the right for SU(3)$_c$ octets.
Panels on the first, second, third and fourth rows are, respectively,  
 for $y=-5/2,-1/2, 1/2$ and $5/2$.
We also show iso-lines of constant $R_{\gamma \gamma}$ (dashed red) and $R_{GG}$ (solid orange).
}
\label{fig:3+2global}
\end{figure*}

Looking at Figs.~\ref{fig:2+1global} and \ref{fig:3+2global} we can   
have a better idea of values of the model parameters $m_1=m_2$ and  
$c$ which may be needed to satisfy experimental requirements in the near 
future, as the situation of Higgs production and diphoton decay get 
settled by the accumulation of LHC data.

In Fig.~\ref{fig:2+1global} we show the allowed region at 68\%, 95\%
and 99\% CL for the 2+1 model parameters that satisfy $S,T$ and $U$
combined.  Panels on the left are for SU(3)$_c$ triplets, on the middle
for SU(3)$_c$ sextets and on the right for SU(3)$_c$ octets.  Panels from 
top to bottom are, respectively, for $y=1,2$ and $3$. We
also show in these plots iso-lines of constant $R_{\gamma \gamma}$
(dashed red) and constant $R_{GG}$ (solid orange).  We see that for
all the SU(3)$_c$ representations it is difficult to explain
$R_{\gamma \gamma} \sim 1.5-2$ with $y=1$ because of the new fermions
contributions to the oblique parameters. 
However an increase of $10-20\%$ is quite feasible. At the same
time $R_{GG} \gsim 0.5-0.75$ is also possible, in particular, for
colored sextets and octets.  For $y=2$ it is possible to explain
$R_{\gamma \gamma}$ as high as $\sim 1.5-2$ and at the same time $R_{GG}
\gsim 0.55$.  This is true for color triplets, sextets and octets.
For $y=3$, it is easy to explain $R_{\gamma \gamma}>1$, nevertheless
for color representations higher than triplet the increase in the
ratio allowed by the oblique parameters might be too much, unless 
one goes to higher masses. Also,
except for the triplet case, it is difficult to accommodate $R_{GG}$
lower than approximately 0.6 or so.  For colored sextets and octets,
it is possible to explain $R_{GG} \sim 0.5$, or even lower, but this
requires  lower masses.

In Fig.~\ref{fig:3+2global} we now show the 68\%, 95\% and 
99\% CL allowed for the 3+2 model parameters that satisfy $S,T$ and $U$ 
combined. Again, panels on the left are for SU(3)$_c$ triplets, on the middle for 
SU(3)$_c$ sextets and on the right for SU(3)$_c$ octets.
Panels on the first, second, third and fourth rows are, respectively,  
 for $y=-5/2,-1/2, 1/2$ and $5/2$.
We observe that independently of the SU(3)$_c$ representation, it is possible to get 
$R_{\gamma \gamma} \sim 1.1-1.5$  ($R_{\gamma \gamma} \sim 1.1-1.25$)
 if $y=-5/2$ ($y=1/2$). However, to get $R_{GG}\sim 0.5-0.75$ one need 
to have $\vert y \vert \lsim 1/2$. For color triplets we have checked
this can also be achieved for $\vert y \vert \lsim  3/2$.

\section{Signals at the LHC}
\label{sec:signals}

We now  come to the discussion of the direct production at the LHC. 
We have calculated the pair production 
cross section of the lightest colored fermion of each model. 
This was done for p-p collisions at the LHC running at a center of
mass energy of 8 TeV, using MadGraph5~\cite{Alwall:2011uj} with
CTEQ6L~\cite{Pumplin:2002vw} parton distribution functions for the
proton and imposing the following loose cuts: $-2.5 < \eta < 2.5$ and $p_T>
20$ GeV for each fermion.
For all models, the main contributions to pair production of the
lightest particle come from gluon-gluon diagrams so that just a marginal 
dependence on the hypercharge (if any) is expected.

\begin{figure*}[tb]
\begin{center}
 \includegraphics[width=0.49\textwidth]{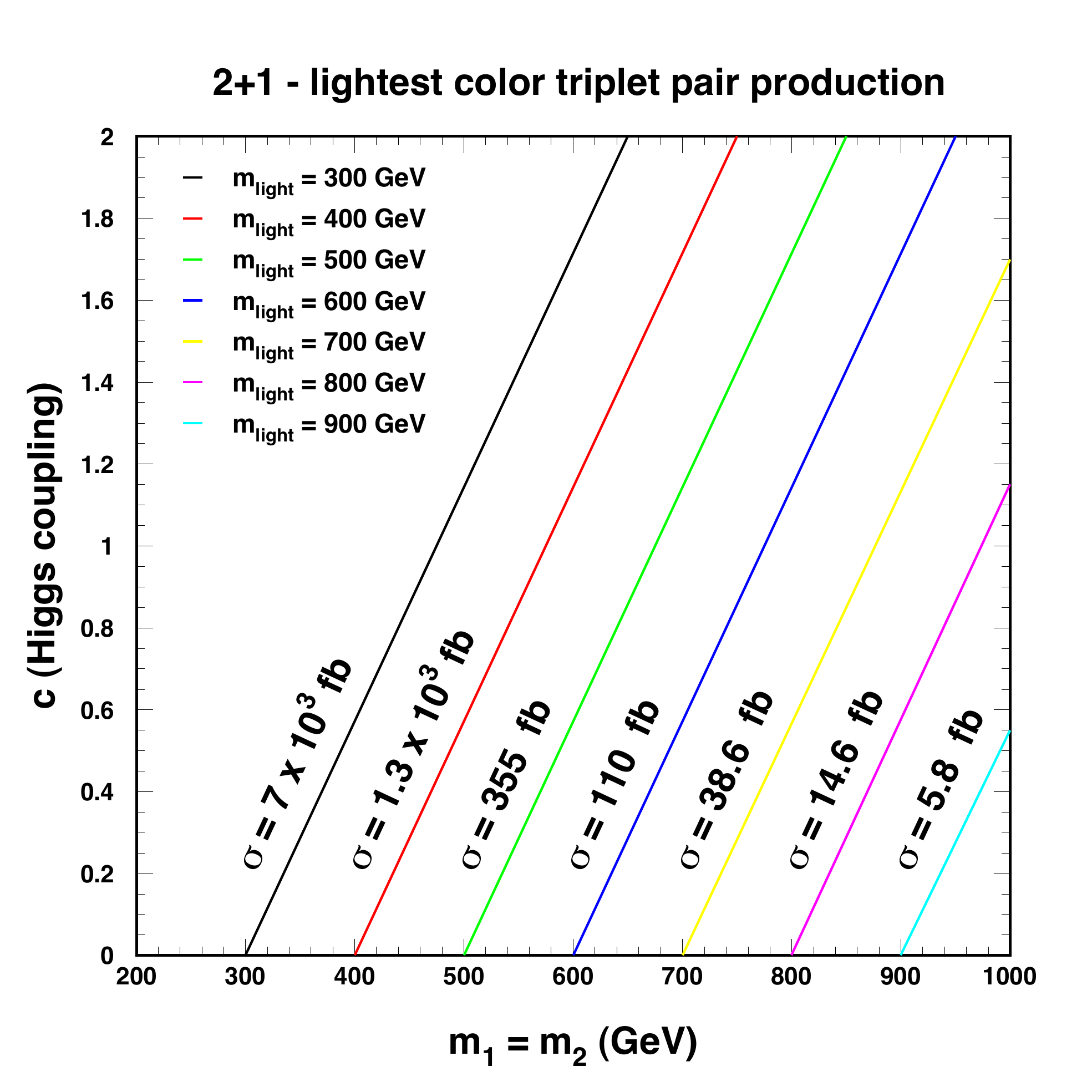}
\vspace{-2mm}
\end{center}
\vspace{-0.1cm}
\caption{Pair production cross section for various masses of the
  lightest SU(3)$_c$ triplet for the $2+1$ model as a function
  of the coupling to the Higgs ($c$) and the vector-like mass $m_1=m_2$.
  Here $y=2$.}
\label{fig:xsec}
\end{figure*}

In Fig.~\ref{fig:xsec} we show the pair production cross section for
the lightest particle of the triplet $2+1$ model with $y=2$ as a
function of $m_1=m_2$ and the coupling to the Higgs ($c$) for various
masses of the lightest particle. We have confirmed that this cross
section changes at the most by a few per cent for other values of $y$
in the 2+1 model.  This is also true for the triplet 3+2 model, so we
do not explicit show a plot for this case.  This cross section,
nevertheless depends on the color representation. 
When going from the triplet to the sextet or octet, 
the gluon-gluon contribution increases by a factor of 20, while 
the $q\bar q$ contribution increases by a factor of $5-6$.
Due to the difference in parton distribution function among gluons and quarks, 
the relative importance between the gluon-gluon and the $q \bar q$ 
contributions to the total cross section will depend on the mass 
of the lightest fermion.

These new fermions can also form superheavy-quarkonia.
Although the calculation performed for the production and decays of 
quarkonium in Ref.~\cite{Barger:1987xg} assumed  a fourth generation of 
heavy chiral quarks, we can use their results to estimate the production and 
decay of the lightest of the heavy bound states.
Production and decay of quarkonium states depend crucially on the 
quarkonium wave function which requires a choice of the interquark potential 
model. There are several possible choices for this potential model.
For our purposes, and for simplicity, we will assume  a Coulomb potential 

\begin{equation} 
V(r) = - \frac{4}{3} \, \frac{\alpha_s(m_Q^2)}{r} \, ,
\label{eq:cornell}
\end{equation}
with the running strong coupling $\alpha_s$ calculated at the scale 
$m_Q$, the mass of the new colored state. Notice that the factor of $4/3$ applies only 
to color triplet, while in general $\frac{4}{3} \rightarrow C_R - \frac{1}{2} C_{\cal{R}}$ \cite{Kats:2012ym}, with 
$C_{R,\;{\cal{R}}}$ the quadratic Casimirs of the constituent fermion and of the bound state respectively. 
Since we will only be interested in order of magnitudes estimates of the relevant quantities, 
we will use the triplet case as representative.

The quarkonium radial wave function squared for the S state 
and the derivative for the P state at the origin ($r=0$) can be 
derived from the potential as
\begin{eqnarray}
\vert R_S(0)\vert^2 = m_Q \, \Big\langle \frac{d V}{dr}\Big \rangle \, , \\
\vert R_P(0)'\vert^2 = \frac{m_Q}{9} \, 
\Big \langle \frac{1}{r^2}\frac{d V}{dr} + 4\frac{(E-V)}{r^3}\Big\rangle \, , 
\label{eq:Rs}
\end{eqnarray}
where $\langle .. \rangle$ denotes the expectation value, 
$E=M_X-2 m_Q$ (with $M_X$ and $m_Q$ the mass of the quarkonium and of the constituent heavy fermion, respectively), so that 
$\vert R_S(0)\vert^2 = 4(2 \alpha_s m_Q/3)^3$ and
$\vert R_P(0)'\vert^2 = (2 \alpha_s m_Q/3)^5/24$.
According to 
\cite{Barger:1987xg} this approximation underestimates 
$\vert R_S(0)\vert^2$ by less than a factor 2,  and 
$\vert R_P'(0)\vert^2$ by a factor 10 or so, but this 
should be enough for our purposes.

We will only consider here the quarkonium bound states with smallest 
spin: the
pseudo-scalar $\eta(0^{-+})$ and the scalar $\chi(0^{++})$. 
They can be directly
produced via the gluon-gluon fusion mechanism.  Their partial widths
into gluons, which dominate by far the decays of the $\eta$ ($\chi$)
up to quarkonium masses of 1 (0.5) TeV, are
\begin{eqnarray}
\Gamma(\eta \to gg) &= &\frac{8 \, \alpha_s^2(m_Q^2)}{3 \, M_\eta^2} \vert R_S(0)\vert^2 \, ,\\
\Gamma(\chi \to gg) &= &\frac{96 \, \alpha_s^2(m_Q^2)}{M_\chi^4} \vert R_P(0)'\vert^2 \, ,
\label{eq:decay1}
\end{eqnarray}
so that $\Gamma(\eta \to gg) \sim {\cal O}$(MeV) and $\Gamma(\chi \to gg) \sim {\cal O}$(keV),  
for $M_\eta=M_\chi \sim 1$ TeV. In both cases we take $m_Q = $ 500 GeV.

The production cross section for a spin-0 quarkonium $X$ via gluon-gluon 
fusion at the LHC can be written in terms of its gluonic decay width
as 
\begin{equation}
\sigma(pp \to gg \to X) = \frac{\pi^2 \tau}{8 \, M_X^3} \,  
\Gamma (X \to gg) \int_\tau^1  \frac{dx}{x} \, g(x,q^2)g(\tau/x,q^2) \, ,
\end{equation}
where $\tau=M_X^2/s$, $s= 4
E_{\rm cm}^2$, $g(x,q^2)$ is the gluon parton distribution function
for the proton.  For $M_\eta = M_\chi \sim $ 1 TeV, we have estimated
$\sigma(pp \to \eta) \sim 0.2$ fb and $\sigma(pp \to \chi) \sim 2
\times 10^{-4}$ fb for the LHC at a center of mass energy of 8 TeV. 
Assuming BR$(\eta,\;\chi \rightarrow gg)\simeq 1$, these results can be compared to 
current experimental limits on dijets signal cross section at Tevatron \cite{Aaltonen:2008dn}, which 
for $M_{\eta,\; \chi} \sim 1$ TeV reads $\sigma_{jj} \lesssim 200$ fb.\\

In general, the
  available experimental limits on new charged colored particles cannot
  be directly applied in a model independent way.
Even so, to have an idea of
possible limits, searches in the LHC for a heavy top-like quark, $t'$,
that decays as $t'\to b\, W$, or $t'\to t \, Z$ place the limit
$M_{t'} \gsim $ 415-557 GeV at 95\% CL~\cite{CMS:2012ab,Rao:2012gf}.
Also searches by CMS for $b'$-quarks that are assumed to decay
exclusively as $b'\to t\, W$ excludes the existence of these quarks
for $M_{b'}$ below 611 GeV~\cite{Chatrchyan:2012yea}.  There are also
limits on stopped long-lived particles at the LHC.  Gluinos and stops
with a lifetime between 10 $\mu s$ to 1000 s and with masses,
respectively, below 640 GeV and 340 GeV are excluded at 95\% CL by LHC
data taken during the 7 TeV proton-proton operation~\cite{:2012yg}.

\section{Conclusion}
\label{sec:conclusions}

We have investigated how new colored vector-like fermions, in the smallest representations 
of $\rm SU(3)_{c} \times SU(2)_{L} \times U(1)_Y$, can contribute to the Higgs to diphoton 
rate and also modify the main Higgs production cross section.
We assume these new particles do not mix with the SM fermions, but since they acquire  
part of their mass from the electroweak symmetry breaking mechanism, they 
mix among themselves.

For concreteness we have analyzed two minimal models with respect to the $\rm SU(2)_L$ 
representation of the fermion fields: the doublet-singlet and the triplet-doublet model. 
In both cases we have supposed the colored fermions to 
be either in the triplet, sextet or octect representation of $\rm SU(3)_c$.
We have established what are the values of the parameters of the models (coupling to 
the Higgs, charges and vector-like masses) that can increase/decrease $\Gamma (h \to \gamma \gamma)$
and $\sigma(gg \to h)$ and still be consistent with the electroweak tests by fitting 
the model to the current bounds on $S,T$ and $U$.

Many of the combined analysis of the available Higgs boson 
experimental results~\cite{Corbett:2012dm,Giardino:2012dp} seem to proclaim   
$h \to \gamma \gamma$ and $gg \to h$ data already point to the need of 
physics beyond the SM. The ATLAS Collaboration~\cite{Atlasconf} also 
points towards $\Gamma(h \to\gamma \gamma) \sim 1.4 
\Gamma(h \to\gamma \gamma)_{\rm SM}$ and $\sigma(gg \to h) \sim 1.3 
\sigma(gg \to h)_{\rm SM}$, but both consistent with the SM at 95 \% CL.

Ultimately, only more LHC data will tell us if the Higgs to diphoton
rate and/or the Higgs production cross section through gluon-gluon
fusion are in fact a door to new physics, or if they can be perfectly
understood in the SM context.  If either one turns out to be away from SM
expectations we provide in this paper some possible scenarios that can
be experimentally testable.

\begin{acknowledgments} 
  \vspace{-0.3cm} 
This work was supported by Funda\c{c}\~ao de Amparo
  \`a Pesquisa do Estado de S\~ao Paulo (FAPESP), Conselho Nacional 
de Desenvolvimento Cient\'ifico e Tecnol\'ogico (CNPq), by the European 
Commission under the contract PITN-GA-2009-237920 and by the Agence National 
de la Recherche under contract ANR 2010 BLANC 0413 01.  
We thank O.J.P. \'Eboli for useful comments and discussions, and L. G. Almeida for 
discussions in the early stages of the project. We also thank Y. Kats for useful comments 
and for pointing out ref.~\cite{Kats:2012ym}.
\end{acknowledgments}


\end{document}